# Probing electronic and magnetic transitions of short periodic nickelate superlattices using synchrotron x-ray

S. Middey,[1, *] Ranjan Kumar Patel,[1] D. Meyers,[2] P. Shafer,[3] M. Kareev,[4]
J. W. Freeland,[5] J.-W. Kim,[5] P. J. Ryan,[5] and J. Chakhalian[4]

[1]*Department of Physics, Indian Institute of Science, Bengaluru 560012, India*

[2]*Department of Physics, Oklahoma State University, Stillwater, Oklahoma 74078, USA*

[3]*Advanced Light Source, Lawrence Berkeley National Laboratory, Berkeley, California 94720, USA*

[4]*Department of Physics and Astronomy, Rutgers University, Piscataway, New Jersey 08854, USA*

[5]*Advanced Photon Source, Argonne National Laboratory, Argonne, Illinois 60439, USA*

**Introduction:** Transition metal based oxide heterostructures exhibit diverse emergent phenomena e.g. two dimensional electron gas, superconductivity, non-collinear magnetic phase, ferroelectricity, polar vortices, topological Hall effect etc., which are absent in the constituent bulk oxides [1–6]. The microscopic understandings of these properties in such nanometer thick materials are extremely challenging. Synchrotron x-ray based techniques such as x-ray diffraction, x-ray absorption spectroscopy (XAS), resonant x-ray scattering (RXS), resonant inelastic x-ray scattering (RIXS), x-ray photoemission spectroscopy, etc. are essential to elucidating the response of lattice, charge, orbital, and spin degrees of freedoms to the heterostructuring [7–14]. As a prototypical case of complex behavior, rare-earth nickelates ($RE$NiO$_3$ with $RE$=La, Pr, Nd, Sm, Eu….Lu) based thin films and heterostructures have been investigated quite extensively in recent years [15–25]. An extensive body of literature about these systems exists and for an overview of the field, we refer the interested readers to the recent reviews 26 and 27. The bulk properties of $RE$NiO$_3$ have been reviewed by Medarde [28] and Catalan [29]. In the present article, we give a brief review that concentrates on the use of synchrotron based techniques to investigate a specific set of EuNiO$_3$/LaNiO$_3$ superlattices, specifically designed to solve a long-standing puzzle about the origin of simultaneous electronic, magnetic and structural transitions of the $RE$NiO$_3$ series [30–32].

Apart from the least distorted member LaNiO$_3$ (space group $R\bar{3}c$), all other members of the $RE$NiO$_3$ bulk series undergo a temperature driven metal-insulator transition (MIT), which is also accompanied by lowering of structural symmetry from orthorhombic (*Pbnm*) to monoclinic (*P*2$_1$/*n*) [33]. Since *P*2$_1$/*n* space group contains two inequivalent Ni sites, the structural transition was also linked to a charge disproportionation (CD) [Ni$^{+3}$ + Ni$^{+3}$ → Ni$^{+3+\delta}$ + Ni$^{+3-\delta}$] transition [33, 34], which has been further corroborated by RXS experiments [35–37]. In recent years, this MIT has been assigned theoretically as a site-selective Mott transition ($d^8\underline{L}+d^8\underline{L} \rightarrow d^8 + d^8\underline{L}^2$, where $\underline{L}$ denotes a hole in the O *p* orbital) primarily due to the very small charge transfer energy of these systems [38–41]. Since NiO$_6$ octahedra with $d^8$ ($S$=1) and $d^8\underline{L}^2$ ($S$=0) configurations have long and short Ni-O bonds, respectively, this insulating state can be also described as a bond disproportionate (BD) phase. The most important outcome of these theoretical works is that the structural symmetry change and the charge ordering on Ni sublattice are not essential to stabilize the insulating phase. Further, Mandal et al. have demonstrated that the BD phase appears when the effective charge transfer energy becomes negative [42]. Surprisingly, signatures of breathing distortion have been also found in the metallic phase of LaNiO$_3$ [43]. On the other hand, there are also reports emphasizing the importance of structural transition [44, 45] and polaron condensation [46]. Fermi surface nesting driven MIT has been also proposed to explain simultaneous electronic and magnetic transitions of NdNiO$_3$ and PrNiO$_3$ [47, 48]. However, experimentally it is extremely difficult to probe if the MIT can be linked with one of those proposed mechanisms as the MIT is always accompanied by another two or three transitions in reality.

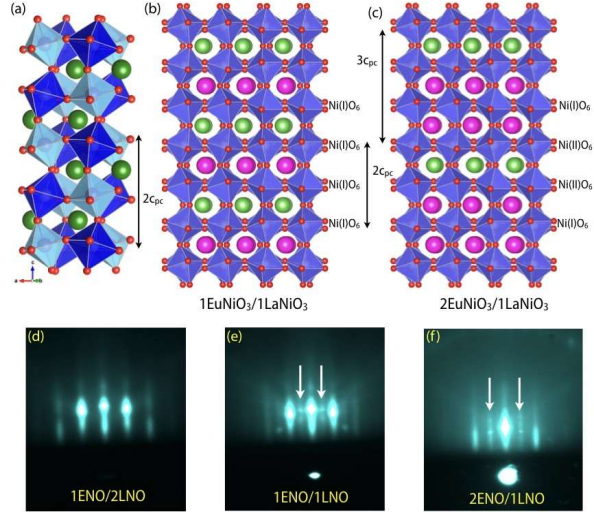

FIG. 1. (a) Crystal structure of bulk $RE$NiO$_3$ in monoclinic, charge ordered phase. (b), (c) Schematic of 1ENO/1LNO and 2ENO/1LNO SLs, respectively. Similar structures were shown in Ref. 30. (d)-(f) RHEED images of 1ENO/2LNO, 1ENO/1LNO and 2ENO/1LNO SLs, respectively along the (1 -1 0)$_{or}$ direction of the NdGaO$_3$ substrate. These RHEED images have been adapted with permission from Ref. 31. Copyright 2018, American Institute of Physics.

In order to address this actively debated issue, we focused on epitaxial growth of short periodic superlattices, consisting of two members of the series: LaNiO$_3$ (LNO) and EuNiO$_3$ (ENO). Since the octahedral rotational patterns of LNO and ENO are different ($a^-a^-a^-$ vs. $a^-a^-c^+$ in Glazer notation [49]), a strong structural competition at the interface between ENO and LNO can be anticipated [50]. Moreover, the ability to grow these materials individually with unit cell (in pseudo-cubic notation) precision allows to mismatch the structural periodicity of the superlattice with the periodicity of a particular ordering pattern. For example, the periodicity of a [2 uc ENO/1 uc LNO] SL (uc= unit cell in pseudo-cubic notation) is $3\times c_{pc}$ along [001]$_{pc}$ (pc=pseudo-cubic), which does not match with the periodicity of the bulk-like checkerboard CO pattern (see Fig. 1(a), (c)). On the other hand, as shown in Fig. 1(b), the periodicity of 1ENO/1LNO SL (=2×$c_{pc}$) is exactly the same as the CO periodicity. In this review article, we summarize our findings of electronic and magnetic structure of such $m$ uc ENO/$n$ uc LNO (mENO/nLNO) SLs using synchrotron based techniques.

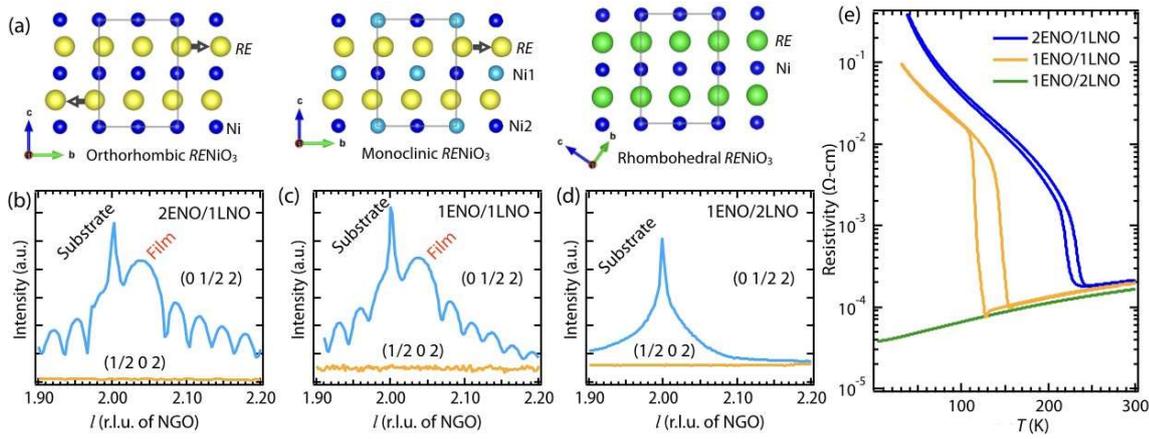

FIG. 2. (a) Schematic of $RE$NiO$_3$ perovskite with different structural symmetry to demonstrate antiparallel displacement of $RE$ ions in orthorhombic and monoclinic symmetry. This feature is absent in the rhombohedral case. (b)-(d) $l$-scan of 2ENO/1LNO and 1ENO/1LNO and 1ENO/2ENO SLs around the (0 1/2 2)$_{pc}$ and (1/2 0 2)$_{pc}$ truncation rods at 300 K. (e) dc resistivity of these films as a function of temperature. These figures have been adapted with permission from Refs. 30 and 31. Copyright 2018, American Physical Society and Copyright 2018, American Institute of Physics.

**Epitaxial growth, structural symmetry, transport measurement:**

These samples were grown on NdGaO$_3$ substrate (pseudo-cubic in-plane lattice constant $a_{sub}$= 3.858Å) by pulsed laser deposition and the layer-by-layer growth of each unit cell of ENO and LNO was confirmed by RHEED (reflection high energy electron diffraction) [30–32]. Apart from the streaking pattern of RHEED image with specular (0 0) and off-specular (±0 1) reflections, additional half order spots (0 ±1/2) were observed for 2ENO/1LNO and 1ENO/1LNO SLs (indicated by arrows in Fig. 1(e)-(f)). The (0 ±1/2) reflections in *RE*NiO$_3$ series arise due to the in-plane doubling of the unit cell with either orthorhombic or monoclinic symmetry [51, 52]. The absence of these half-order spots in 1ENO/2LNO SL (Fig. 1(d)) strongly suggests rhombohedral symmetry of the sample, similar to bulk LNO. The strong structural competition between the ENO and LNO layers were also observed during the growth as the deposition of second LNO layers in each period of 1ENO/2LNO SL results in the disappearance of the half order spots, which reappear after the growth on ENO layer in the next period. The symmetries of these SLs were further checked by synchrotron x-ray diffraction. As illustrated in Fig. 2(a), both orthorhombic and monoclinic *AB*O$_3$ perovskites have an antiparallel displacement of *A*-sites whereas rhombohedral *AB*O$_3$ does not show this kind of behavior. The presence of such antiparallel displacement can be verified by examining half order x-ray diffraction peaks with the index (odd/2 even/2 even/2) [53]. Our findings of the (0 1/2 2)$_{pc}$ diffraction peaks for both the substrate and the film (Fig. 2(b), (c)) for 2ENO/1LNO and 1ENO/1LNO SLs clearly establish orthorhombic/monoclinic symmetry. Moreover, the absence of (1/2 0 2)$_{pc}$ peak for the film and substrate confirm that the film is a single domain and has the same in-plane orientation as the substrate. The rhombohedral symmetry of 1ENO/2LNO film is authenticated by the absence of both (0 1/2 2)$_{pc}$ and (1/2 0 2)$_{pc}$ reflections, shown in Fig. 2(d).

Before discussing the electronic behaviors of these SLs, we recap that it is well established in the literature that the bulk rhombohedral LNO is a paramagnetic metal down to very low temperature. However, a recent claim of an antiferromagnetic metallic phase of LNO in single crystalline form has led to a debate about the actual nature of magnetic ground states [54–56]. On the other hand, bulk ENO undergoes MIT around 460 K and another transition from paramagnetic insulating (PI) phase to antiferromagnetic insulating (AFI) around 200 K [28, 29]. In accordance with its' rhombohedral structure, 2LNO/1ENO SL remains metallic down to the lowest temperature of measurement of ∼ 2 K. On the other hand, both 2ENO/1LNO and 1ENO/1LNO SLs exhibit temperature-driven first order MIT with strong thermal hysteresis, as shown in Fig. 2(e). While the resistivity values in the metallic phases of the SLs are very similar, the transition temperature strongly depends on layer numbers $m$, and $n$.

**Soft X-ray scattering to probe magnetic ordering:**

The magnetic structure of polycrystalline NdNiO$_3$ and PrNiO$_3$ samples was initially solved by the neutron diffraction experiment [57]. The unusual $E'$ antiferromagnetic spin configuration is characterized by a magnetic wave vector (1/2 0 1/2)$_{or}$ ($\equiv$(1/4 1/4 1/4)$_{pc}$), which can be viewed as a spin sequence of either ↑↑↓↓ or ↑→↓← along [1 1 1]$_{pc}$. Resonant soft x-ray scattering experiment on single crystalline NdNiO$_3$ films by Scagnolli et al. further confirmed the non-collinear spin arrangements. In these scattering experiments, the presence of a (1/4 1/4 1/4)$_{pc}$ diffraction peak is examined by x-ray photons with energy close to the Ni $L_3$ edge [58]. This technique has been used extensively to investigate the magnetic ordering of various $RE$NiO$_3$ based thin films, superlattices, and heterostructures, etc [59–65].

Based on the RXS measurement, the antiferromagnetic transition temperature is found to be 220±5 K for 2ENO/1LNO from (Fig. 3(a)-(b)). An energy scan across the Ni $L_{3,2}$ edges for the (1/4 1/4 1/4)$_{pc}$ Bragg peak is compared (Fig. 3(c)) with simultaneously collected fluorescence background, which represents XAS in fluorescence mode. The resonant enhancement is very strong around 852.8 eV, where XAS also exhibits a sharp peak. Recent RIXS measurements and cluster calculations have attributed this XAS feature with a $d^8$ state [66, 67]. Moreover, Ni $L_{3,2}$ XMCD (x-ray magnetic circular dichroism) of the 2ENO/1LNO SL looks very similar to the $S$=1 Ni XMCD of Ca$_2$NiOsO$_6$ [68]. Furthermore, the magnetic scattering derived from the XMCD spectrum by a Kramers-Kronig (K-K) transformation matches well with the line shape and line position of the scattered intensity, observed in the RXS measurement [30, 69, 70]. All of these studies confirm the presence of an $S$=1 state in 2ENO/1LNO SL, which is contrary to the ionic picture, where $S$=1/2 is expected. This result in turn implies that the sequence of the spin arrangement in (1 1 1)$_{pc}$ planes is ↑ 0 ↓ 0. In addition, the RXS measurements of 1ENO/1LNO SL determined $T_N \sim$ 150 K [30] and the energy scan appears very similar to 2ENO/1LNO SL, shown in Fig. 3(c)).

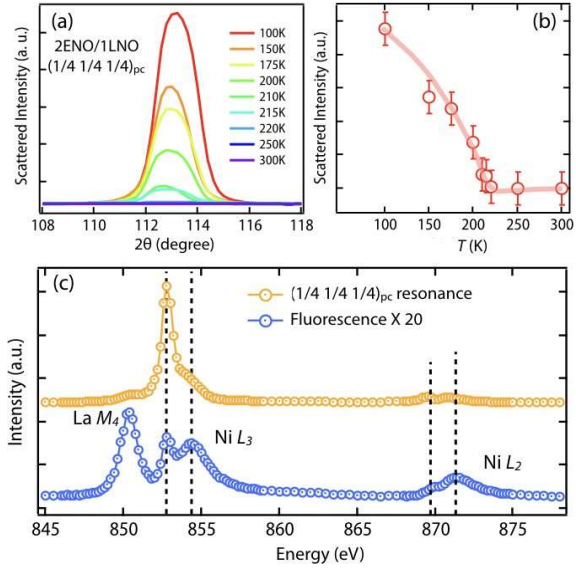

FIG. 3. (a) The resonant x-ray scattering measurement at $(1/4\ 1/4\ 1/4)_{pc}$ peak of 2ENO/1LNO SL. at different temperatures. (b) Integrated area of this peak as a function of temperature shows that the AFM transition temperature is around 220 K. This has been adapted with permission from Ref. 30. Copyright 2018, American Physical Society. (c) Resonance and fluorescence scans of the Ni $L_{3,2}$ edges at 50 K for the 2ENO/1LNO SL. For visual clarity, fluorescence data have been multiplied by 20 and has been shifted vertically.

**Hard X-ray scattering to probe charge ordering:**

The checkerboard type charge ordering (CO) in a 50 nm thick NdNiO$_3$ film was demonstrated using resonant hard x-ray scattering at the Ni $K$ edge by Staub et al [35]. These experiments are focused on $(h\ 0\ l)_{or}$ and $(0\ k\ l)_{or}$ reflections with $h$, $k$, $l$ being odd integers, as these peaks have no contribution from Ni in *Pbnm* symmetry and inequivalent Ni sites in the monoclinic structure give rise to a strong energy dependence for these particular Bragg reflections. The observation of CO was further corroborated by polarization dependent analysis and azimuthal scans [36, 37]. However, it has been argued in very recent years that the Ni resonance features can be obtained by BD also, without any CO on the Ni sublattice [71, 72]. Similar experiments on 6 nm thin NdNiO$_3$ films did not find any signature of Ni resonance, and has been interpreted as a suppression of CO by epitaxial strain [73, 74].

The 1ENO/1LNO SL, which exhibits simultaneous MIT and magnetic transition around 150 K, also showed sharp resonance features in the insulating phase (see Fig. 4(a)). As shown in Fig. 4(b), the temperature dependence plot of the resonant intensity demonstrates a CO transition around 150 K. The off-resonant intensity, which depends on the crystal structure, also shows a transition at the same temperature. This result confirms that the structural changes and the CO are cooperative in nature in this system. In contrast, the resonance line shape of the 2ENO/1LNO SL, and resonance and off-resonance intensity did not show any significant modulation across the $T_{MIT}$ ~ 245 K. This observation establishes that a metal-insulator transition can be obtained without any bulk-like charge ordering and structural symmetry change. Moreover, the strong resonance features in the metallic-phase of 2ENO/1LNO SL is evidence of a rare monoclinic metal phase. The observation of $S=1$ state of Ni, together with negligible orbital polarization also support the site-selective Mott transition scenario in this artificial material [38–41].

**Effect of epitaxial strain:**

So far, all of the results discussed in this paper are for ENO/LNO heterostructures grown on $NdGaO_3$ substrates. In order to investigate the effect of the epitaxial strain on the electronic and magnetic transition, 1ENO/1LNO SLs were also grown on $DyScO_3$ (in-plane pseudo-cubic lattice constant $a_{sub}$=3.955Å), $LaAlO_3$ ($a_{sub}$=3.794Å), and $YAlO_3$ ($a_{sub}$=3.692Å) substrates. Interestingly, the electronic transition temperature does not change appreciably with the increase of the tensile strain. Most importantly, both the MIT and antiferromagnetic ordering transition are completely suppressed under the application of compressive strain (Fig. 5(a)). Similar suppression of MIT has been also reported for $PrNiO_3$, $NdNiO_3$, $SmNiO_3$ and $EuNiO_3$ thin films [21, 59, 61, 62, 75, 76].

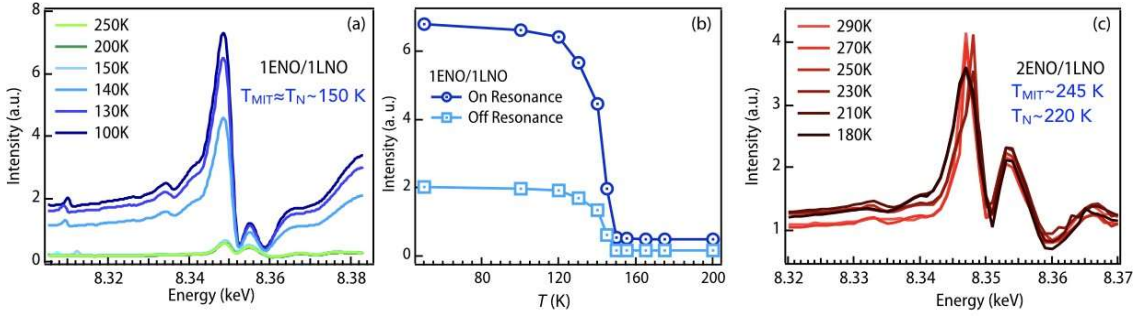

FIG. 4. $(0\ 0\ 1)_{or} \equiv (-1/2\ 1/2\ 1/2)_{pc}$ resonance energy scan for (a) 1ENO/1LNO and (c) 2ENO/1LNO SLs at different temperature. (b) Temperature dependence of resonance intensity ($E_{res}$=8.349 keV) and offresonant intensity ($E_{off-res}$=8.33 keV) of 1ENO/1LNO SL. (a) and (c) have been adapted with permission from Ref. 30. Copyright 2018, American Physical Society.

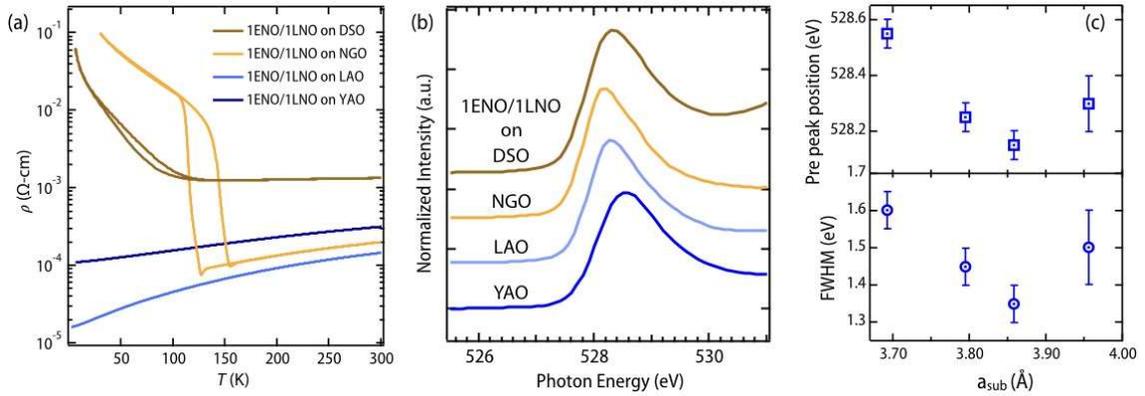

FIG. 5. (a) Temperature dependent dc resistivity, (b) pre peak around 528.5 eV of O-K edge XAS at 300 K for 1ENO/1LNO SLs on different substrates. The higher resistivity at 300 K of the SL grown on DyScO$_3$ signifies the presence of a small amount of oxygen vacancies. (c) Position and FWHM of the pre peak as a function of in-plane lattice constant. The figure has been adapted with permission from Ref. 32. Copyright 2018, American Physical Society.

To understand this strain effect, O *K*-edge resonant x-ray absorption spectra were measured at room temperature on these SLs [77]. In this technique, a pre peak is observed around 528 eV due to the $d^8\underline{L} \rightarrow \underline{c}d^8$ transition (here $\underline{c}$ is a hole in the oxygen 1*s* core state) and the intensity, position, and width of this pre peak depend on the Ni-O covalency (Fig. 5b)) [18, 59]. As the strain becomes more compressive, the pre peak shifts towards higher energy (Fig. 5(c)). This signifies a decrease in charge transfer energy ($\Delta$) and is related to the strong modulation of relative Madelung potential between Ni and O by epitaxial strain [18, 59, 78]. Moreover, the increase of FWHM of the pre peak with the compressive strain implies the increase of Ni-O covalency. Thus, strain induced modulation of both $\Delta$ and hybridization energy are responsible for the suppression of the antiferromagnetic insulating phase.

**Summary:**

In this brief review article, we discuss about the results of synchrotron based x-ray diffraction, x-ray absorption spectroscopy, and resonant (soft and hard) x-ray scattering experiments on a series of short periodic EuNiO$_3$/LaNiO$_3$ superlattices. These experiments demonstrate that the 2EuNiO$_3$/1LaNiO$_3$ SL exhibits a temperature driven metal-insulator transition without any structural symmetry change and charge ordering, solving a long-standing puzzle about the origin of simultaneous transition in *RE*NiO$_3$ series. Moreover, the results of these experiments emphasize the unique insights obtained by combining synchrotron based techniques to understand multiple electronic, spin, and lattice order parameters in nanometer thin artificial quantum materials.

**Acknowledgements:**

SM is funded by a DST Nanomission grant (DST/NM/NS/2018/246) and a SERB Early Career Research Award (ECR/2018/001512). J.C. is supported by the Gordon and Betty Moore Foundation EPiQS Initiative through Grant No. GBMF4534. M. Kareev was supported by the Department of Energy under grant DE-SC0012375. This research used resources of the Advanced Photon Source, a U.S. Department of Energy Office of Science User Facility operated by Argonne National Laboratory under Contract No. DE-AC02-06CH11357. This research used resources of

———

*smiddey@iisc.ac.in

1. H. Y. Hwang et al., *Nature Mater.* **11**, 103 (2012).
2. J. Chakhalian et al., *Rev. Mod. Phys.* **86**, 1189 (2014).
3. P. Zubko et al., *Annual Review of Condensed Matter Physics* **2**, 141 (2011).
4. S. Stemmer et al., Annual Review of Materials Research **44**, 151 (2014).
5. J. Matsuno et al., *Science advances* **2**, e1600304 (2016).
6. A. Yadav et al., *Nature* **530**, 198 (2016).
7. F. De Groot, *Chemical Reviews* **101**, 1779 (2001).
8. J. Fink et al. *Reports on Progress in Physics* **76**, 056502 (2013).
9. B. Pal et al., *Journal of Electron Spectroscopy and Related Phenomena* **200**, 332 (2015).
10. J. Chakhalian et al., *Science* **318**, 1114 (2007).
11. J. Chakhalian et al., *Nat Phys* **2**, 244 (2006).
12. D. D. Fong et al., *Annu. Rev. Mater. Res.* **36**, 431 (2006).
13. A. Frano et al., *Nature materials* **15**, 831 (2016).
14. D. Meyers et al., *Phys. Rev. Lett.* **121**, 236802 (2018).
15. J. c. v. Chaloupka et al., *Phys. Rev. Lett.* **100**, 016404 (2008).
16. P. Hansmann et al., *Phys. Rev. Lett.* **103**, 016401 (2009).
17. A. V. Boris et al., *Science* **332**, 937 (2011),
18. J. Chakhalian et al., *Phys. Rev. Lett.* **107**, 116805 (2011).
19. R. Scherwitzl et al., *Phys. Rev. Lett.* **106**, 246403 (2011).
20. A. M. Kaiser et al., *Phys. Rev. Lett. **107**, 116402 (2011)*.
21. D. Meyers et al., *Phys. Rev. B* **88**, 075116 (2013).
22. R. Jaramillo et al., *Nat Phys* **10**, 304 (2014).
23. P. D. C. King, et al., *Nat Nano* **9**, 443 (2014).
24. A. S. Disa et al., *Phys. Rev. Lett.* **114**, 026801 (2015).
25. S. Middey et al., *Phys. Rev. Lett.* **116**, 056801 (2016).
26. S. Middey et al., Annual Review of Materials Research **46**, 305 (2016).
27. S. Catalano et al., *Reports on Progress in Physics* **81**, 046501 (2018).
28. M. L. Medarde, *Journal of Physics: Condensed Matter* **9**, 1679 (1997).

29. G. Catalan, *Phase Transitions* **81**, 729 (2008).

30. S. Middey et al., *Phys. Rev. Lett.* **120**, 156801 (2018).

31. S. Middey et al., *Applied Physics Letters* **113**, 081602 (2018).

32. S. Middey et al., *Phys. Rev. B* **98**, 045115 (2018).

33. J. A. Alonso et al., *Phys. Rev. Lett.* **82**, 3871 (1999).

34. I. I. Mazin et al., *Phys. Rev. Lett.* **98**, 176406 (2007).

35. U. Staub et al., *Phys. Rev. Lett.* **88**, 126402 (2002).

36. J. E. Lorenzo et al., *Phys. Rev. B* **71**, 045128 (2005).

37. V. Scagnoli et al., *Phys. Rev. B* **72**, 155111 (2005).

38. T. Mizokawa et al., *Phys. Rev. B* **61**, 11263 (2000).

39. H. Park et al., *Phys. Rev. Lett.* **109**, 156402 (2012).

40. S. Johnston et al., *Phys. Rev. Lett.* **112**, 106404 (2014).

41. A. Subedi et al., *Phys. Rev. B* **91**, 075128 (2015).

42. B. Mandal et al., arXiv preprint arXiv:1701.06819 (2017).

43. V. L. Karner et al., Phys. Rev. B **100**, 165109 (2019).

44. A. Mercy et al., *Nature Communications* **8**, 1677 (2017).

45. O. E. Peil et al., *Phys. Rev. B* **99**, 245127 (2019).

46. J. Shamblin et al., *Nature Communications* **9**, 86 (2018).

47. S. Lee et al., *Phys. Rev. B* **84**, 165119 (2011).

48. S. Lee et al., *Phys. Rev. Lett.* **106**, 016405 (2011).

49. A. M. Glazer, *Acta Crystallographica Section B* **28**, 3384 (1972).

50. I. C. Tung et al., *Phys. Rev. B* **88**, 205112 (2013).

51. D. Meyers et al., *J. Phys. D: Appl. Phys.* **46**, 385303 (2013).

52. S. K. Ojha et al., *Phys. Rev. B* **99**, 235153 (2019).

53. G. A. Ravi et al., *J. Am. Ceram. Soc.* **90**, 3947 (2007).

54. H. Guo et al., *Nature communications* **9**, 43 (2018).

55. J. Zhang et al., *Crystal Growth & Design* **17**, 2730 (2017).

56. B.-X. Wang, et al., *Physical Review Materials* **2**, 064404 (2018).
57. J. L. García-Muñoz et al., *Phys. Rev. B* **50**, 978 (1994).
58. V. Scagnoli et al., *Phys. Rev. B* **73**, 100409 (2006).
59. J. Liu et al., *Nat Commun* **4**, 2714 (2013).
60. A. Frano et al., *Phys. Rev. Lett.* **111**, 106804 (2013).
61. M. Hepting et al., *Phys. Rev. Lett.* **113**, 227206 (2014).
62. S. Catalano et al., *APL Materials* **2**, 116110 (2014).
63. S. Catalano et al., *APL materials* **3**, 062506 (2015).
64. D. Meyers et al., *Phys. Rev. B* **92**, 235126 (2015).
65. M. Hepting et al., *Nature Physics* **14**, 1097 (2018).
66. V. Bisogni et al., *Nature Communications* **7**, 13017 (2016).
67. R. J. Green et al., *Phys. Rev. B* **94**, 195127 (2016).
68. R. Morrow et al., *Chemistry of Materials* **28**, 3666 (2016).
69. J. F. Peters et al., *Phys. Rev. B* **70**, 224417 (2004).
70. C. Schüßler-Langeheine et al., *Phys. Rev. Lett.* **95**, 156402 (2005).
71. K. Haule et al., *Scientific Reports* **7**, 2045 (2017).
72. Y. Lu et al., *Phys. Rev. B* **93**, 165121 (2016).
73. M. H. Upton, et al., *Phys. Rev. Lett.* **115**, 036401 (2015).
74. D. Meyers, et al., *Scientific Reports* **6**, 27934 (2016).
75. E. Mikheev, et al., *Science Advances* **1**, e1500797 (2015),
76. F. Y. Bruno et al., *Phys. Rev. B* **88**, 195108 (2013).
77. F. M. F. de Groot et al., *Phys. Rev. B* **40**, 5715 (1989).
78. M. Imada et al., *Rev. Mod. Phys.* **70**, 1039 (1998).